# Superconductor-semiconductor hybrid cavity quantum electrodynamics

Guido Burkard[1], Michael J. Gullans[2], Xiao Mi[2,*], and Jason R. Petta[2]

1. Department of Physics, University of Konstanz, D-78457 Konstanz, Germany
2. Department of Physics, Princeton University, Princeton, New Jersey 08544, USA

* Present address: Google Inc., Santa Barbara, California 93117, USA

Abstract | Light-matter interactions at the single particle level have generally been explored in the context of atomic, molecular, and optical physics. Recent advances motivated by quantum information science have made it possible to explore coherent interactions between photons trapped in superconducting cavities and superconducting qubits. Spins in semiconductors can have exceptionally long spin coherence times and can be isolated in silicon, the workhorse material of the semiconductor microelectronic industry. Here, we review recent advances in hybrid "super-semi" quantum systems that coherently couple superconducting cavities to semiconductor quantum dots. We first present an overview of the underlying physics that governs the behavior of superconducting cavities, semiconductor quantum dots, and their modes of interaction. We then survey experimental progress in the field, focusing on recent demonstrations of cavity quantum electrodynamics in the strong coupling regime with a single charge and a single spin. Finally, we broadly discuss promising avenues of future research.

A remarkable experimental achievement of the 1980's and 1990's was to create minimalistic hybrid systems consisting of only one single atom, which exists in one of two states, and interacts with individual photons in a cavity[1-4]. This field of research, termed cavity QED, showed that it is possible to create a quantum superposition of light and matter[5]. More complex systems such as superconductors and semiconductors can themselves be building blocks in hybrid systems on a larger scale. In the early 2000's, cavity QED was realized in condensed matter systems using self-assembled quantum dots confined in photonic cavities[6-8] and by placing a superconducting qubit inside of a microwave cavity[9,10]. In these experiments the atom that is conventionally used in atomic cavity QED is replaced with a quantum device that has discrete energy levels whose energy separation can be matched to the energy of a cavity photon. Around the same time, it was conjectured that cavity QED could be performed using individual electrons trapped in gate defined semiconductor quantum dots, using either charge or spin degrees of freedom to mimic the states of an atom[11-13].

There are a number of motivations for examining cavity QED in the context of condensed matter systems, many of which are grounded in the rapidly growing field of quantum information science. On the heels of the discovery that a superconducting circuit could be coherently coupled to microwave photons[10] were two experiments showing that two spatially separated superconducting qubits could be coupled via a cavity[14,15]. The then nascent subject of circuit QED has now expanded into a field of its own. Prominent advances include demonstrations of multiqubit entanglement[16-19], readout of quantum states[20-22], the generation of non-classical light[23-25], the development of error correction based on Schrodinger cat states[26,27], quantum feedback[28,29] and measurements of quantum trajectories[30]. Fundamentally, experiments involving superconducting quantum devices take advantage of a macroscopic superconducting condensate that is protected by an energy gap $\Delta$ (e.g. $\Delta \sim 175$ μeV in Al, roughly 20 times larger than the thermal energy 100 mK ~ 8 μeV in typical experiments). This begs the question: can cavity QED physics be explored with single charges and spins in semiconductor devices, where such protection is absent? The prospect of cavity QED with a single spin is especially intriguing, as spin coherence times can exceed seconds in some solid state systems[31-33].

In this Review, we describe dramatic developments in the area of "hybrid" circuit QED, where gate defined quantum dots are coupled to superconducting cavities in a "super-semi" device architecture. Recent demonstrations of strong coupling physics with single charges and spins confined in semiconductor quantum dots make this a timely topic to review[34-38]. We begin by laying the theoretical groundwork for the experiments, with a description of the superconducting cavity, the "artificial atom" which in most experiments consists of a semiconductor double quantum dot (DQD), their modes of interaction, and the figures of merit that succinctly describe the quantum coherence of the system. We then survey experiments involving the charge degree of freedom, which interacts with the cavity electric field through the electric dipole interaction[34, 35]. A combination of electric dipole coupling and spin-orbit coupling enables coherent spin-photon interactions, which we review next[36, 37]. Lastly, we give several examples illustrating how semiconductor circuit QED could impact fundamental science and engineering in diverse areas ranging from topological physics to surface microscopy and quantum technology.

## 1. Cavity QED with double quantum dots

At a basic level, a typical cavity QED system (FIG 1a) consists of just two components: a cavity that supports a well-defined photon mode at a cavity resonance frequency $f_c$, and a two-level quantum system with a transition energy $E_{|1\rangle} - E_{|0\rangle}$ that is closely matched to the energy of a photon trapped in the cavity $hf_c$, where $h$ is Planck's constant. Here $E_{|0\rangle}$ ($E_{|1\rangle}$) is the ground state (first excited state) energy. The first atomic physics demonstrations of cavity QED used microwave transitions between Rydberg states of single cesium atoms[39, 40] These results were eventually extended to the visible spectral range[41]. Cavity QED can also be implemented using a wide variety of solid-state systems, as illustrated with some examples in FIG 1b. Color centers in diamond such as nitrogen vacancy (NV) centers have spin-full ground states and narrow, spin-selective microwave and optical transitions, which allows the realization of cavity QED using integrated photonic structures[42-46]. Using nanofabrication techniques, it is possible to build mesoscopic semiconducting and superconducting devices that are quantum coherent. Semiconductor DQDs can be used to isolate single electrons, where the charge degree of freedom can be controlled with electric fields[47-49] and the spin degree of freedom with magnetic fields[50] and the exchange interaction[51]. Superconducting circuits combine a capacitance $C$ with a Josephson inductance $L_J$ to create a quantum system with an anharmonic energy level spectrum[52-57]. Cavity QED experiments involving superconducting quantum devices are reviewed in REFS[58, 59]. We focus here on experiments involving semiconductor DQDs[60, 61], as they are electrically tunable and open the door to cavity QED using long-lived spin states (see FIG 2a).

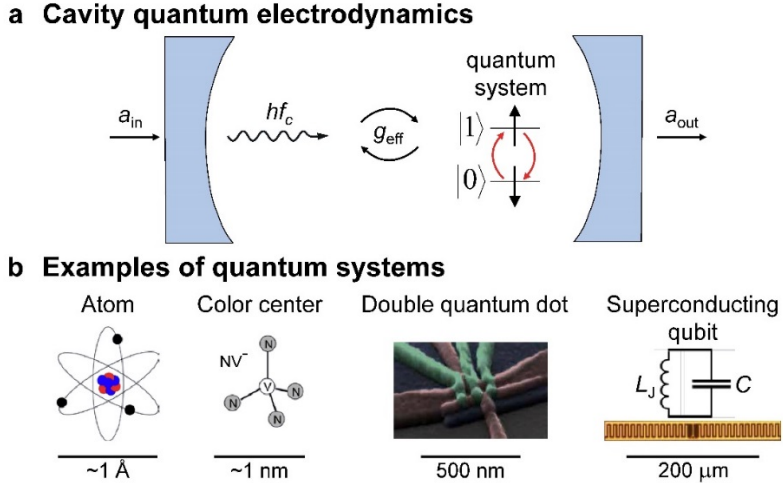

Figure 1 | **Cavity quantum electrodynamics. a** | Cavity quantum electrodynamics (cavity QED) explores the interaction between light and matter at the single particle level. In general, a quantum system with excited (ground) state energy $E_{|1\rangle}$ ($E_{|0\rangle}$) is placed inside of a high quality factor cavity that traps photons of energy $hf_c$. Cavity losses are described by a cavity decay rate $\kappa$ (not shown). The quantum system interacts with the electromagnetic field of the cavity and the interaction is characterized by a coupling frequency $g_{\text{eff}}$. In the dispersive regime, $E_{|1\rangle} - E_{|0\rangle} - hf_c \gg \hbar g_{\text{eff}}$, and the quantum system weakly interacts with the cavity field. Readout can be performed in the dispersive regime by driving the cavity with a weak input field $a_{\text{in}}$ and measuring the transmission through the cavity $a_{\text{out}}/a_{\text{in}}$. In the resonant regime, $E_{|1\rangle} - E_{|0\rangle} \approx hf_c$. Here the quantum system hybridizes with the photonic mode, forming a superposition state of light and matter. **b** Cavity QED has been implemented with many different quantum systems. Early work in atomic physics focused on atoms, but the field has branched out to include color centers, semiconductor DQDs, and superconducting qubits.

**Box 1: Double quantum dot**

A quantum dot (QD) is a nanoscale object that confines an electron in all three spatial dimensions[60-62]. Single quantum dots are described by the electrostatic charging energy $E_c = e^2/2C$, which is the energy cost to add or remove an electron from the system. Here $C \approx 4\pi\epsilon_r\epsilon_o a_o$ denotes the capacitance of the QD and $e$ the elementary charge of an electron, where $\epsilon_r$ is the (relative) dielectric constant, $\varepsilon_0$ is the permittivity of free space, and $a_0$ is the radius of the quantum dot. The orbital "particle-in-a-box" energy scale is governed by $E_{\text{orb}} \sim \hbar^2/m^* a_0^2$, where $\hbar = h/2\pi$ is the reduced Planck constant and $m^*$ is the effective mass of the electron. Both of these energy scales are set by the physical dimensions of the dot ($a_0$) and materials parameters ($m^*$ and $\epsilon_r$), and are therefore difficult to change in-situ. Fortunately, it is possible to make an artificial molecule by placing two quantum dots in proximity to each other and forming a DQD. In semiconductor DQDs, the energy level separation $\varepsilon$ and the interdot tunneling rate $t_c$ can be electrically tuned (FIG 2a).

A DQD containing a single electron can be viewed as a charge qubit; a voltage-tunable double well potential containing a single charge, as illustrated in FIG 2b [47-49, 63]. The charge physics of a DQD is described by the Hamiltonian

$$H_0 = \begin{pmatrix} \varepsilon/2 & t_c \\ t_c & -\varepsilon/2 \end{pmatrix}$$

which is written in the basis $|L\rangle = |(1,0)\rangle, |R\rangle = |(0,1)\rangle$, where $|(N_L, N_R)\rangle$ denotes a DQD charge state with $N_L$ ($N_R$) electrons occupying the left (right) dot[47-49]. In other words, $|L\rangle$ and $|R\rangle$ describe two charge states where the electron is in the left or right dot, respectively. If the left and right dot energy levels are aligned ($\varepsilon=0$) then the $|L\rangle$ and $|R\rangle$ states hybridize to form molecular bonding and antibonding states $|\pm\rangle \propto |L\rangle \pm |R\rangle$, while at large detuning ($|\varepsilon| \gg t_c$) the states $|L\rangle$ and $|R\rangle$ are essentially unperturbed by tunneling. The energy of the two levels is plotted as a function of level detuning $\varepsilon$ in FIG 2c. The states $|+\rangle$ and $|-\rangle$ allow for electric dipole transitions around $\varepsilon \approx 0$ and play the role of the atomic levels for cavity QED[1]. To take into account the fact that an electron is endowed with a spin-½ degree of freedom, we extend our basis to $|(\uparrow,0)\rangle, |(\downarrow,0)\rangle, |(0,\uparrow)\rangle, |(0,\downarrow)\rangle$, where the arrow indicates the spin state of the electron[64].

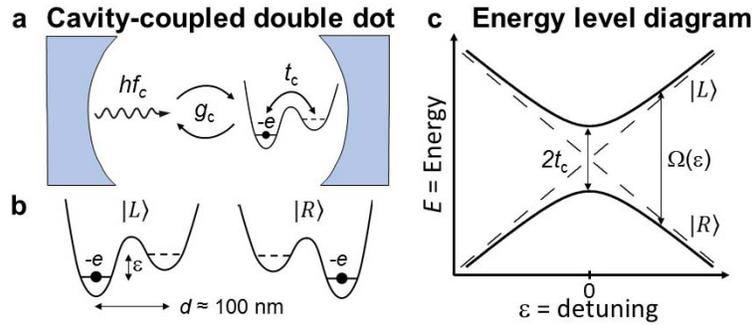

Figure 2 | **Physics of the cavity-coupled double quantum dot. a** | Electric dipole interactions couple a single electron trapped in the DQD to single cavity photon with a strength described by the charge-photon coupling rate $g_c$. **b** The two charge states of the DQD, $|L\rangle$ or $|R\rangle$, are defined by the presence of one electron on the left or right dot, respectively. The interdot spacing is typically on the order of $d = 100$ nm, which leads to a substantial electric dipole moment. **c** Energy level diagram of the DQD charge states. Both $\varepsilon$ and $t_c$ are electrically tunable with gate voltages. As the detuning parameter is increased from zero, the effective dipole moment is reduced by a factor $\frac{2t_c}{\Omega(\varepsilon)}$, where $\Omega = E_{|L\rangle} - E_{|R\rangle} = \sqrt{\varepsilon^2 + 4t_c^2}$ and becomes negligible for large detunings.

*Charge-photon interaction*

A single electron trapped in a DQD forms a fully tunable two-level system (see Box 1). The basic physics of charge-photon coupling in a DQD system is illustrated in FIG. 2a. Electric dipole interactions couple the electron trapped in the DQD to the cavity photon with a strength described by the charge-photon coupling rate $g_c$. The interdot spacing is typically on the order of $d = 100$ nm, which leads to an electric dipole moment $ed$ that is about 1000x larger than the dipole moment of a single atom. The coupling rate $g_c$ is given by the product of this dipole moment with the vacuum (rms) electric field $E_0$ of the cavity. Near zero detuning ($\varepsilon$=0) the charge states are strongly hybridized, leading to the maximum in the charge-photon coupling rate. Away from zero detuning, the $|L\rangle$-$|R\rangle$ charge states are weakly admixed, which reduces the effective dipole moment by a factor $\frac{2t_c}{\Omega(\varepsilon)}$, where $\Omega(\varepsilon) = E_{|L\rangle} - E_{|R\rangle} = \sqrt{\varepsilon^2 + 4t_c^2}$.

Fabry-Perot cavities are typically employed in atomic physics, where optical photons are trapped[41]. For the much larger quantum dot devices, typical energy scales are on the order of 20-40 μeV and it becomes convenient to use superconducting resonators to trap microwave frequency photons (1 GHz ~ 4.2 μeV). A cavity is never perfect and there can be internal losses, described by a decay rate $\kappa_{int}$, and losses through the ports of the cavity, $\kappa_1$ and $\kappa_2$. Cavity QED systems can be probed by measuring the transmission through (or reflection off of) the cavity. For example, in FIG 1a, port 1 is being driven by a weak input field $a_{in}$ and the signal exiting port 2 of the cavity $a_{out}$ is being measured.

When a DQD is placed inside a superconducting microwave resonator, the electric field $E_{res}$ inside the resonator tilts the energy landscape and the difference $\varepsilon$ between the left and right energy levels becomes $\varepsilon + eE_{res}d$. Here, since $d$ is much smaller than the wavelength of the electromagnetic waves inside the resonator, we can apply the electric dipole approximation where $E_{res}$ is constant within the entire volume of the DQD. The quantized electric field operator can be expressed in terms of creation and annihilation operators $a$ and $a^\dagger$ of the electromagnetic field mode inside the resonator (these are equivalent to the ladder operators of the quantum harmonic oscillator), as $E_{res} = E_0(a + a^\dagger)$ where $E_0$ is the vacuum amplitude of the electric field. Taken together, the coupling of the charge qubit to the resonator mode is described with the Hamiltonian $H = H_0 + H_{int}$ with $H_{int} = g_c(a + a^\dagger)\tau_z$ in units where $\hbar$=1, with the charge-cavity coupling $g_c = eE_0 d$ and the quantum operator $\tau_z$ defined via $\tau_z|(1,0)\rangle = |(1,0)\rangle$ and $\tau_z|(0,1)\rangle = -|(0,1)\rangle$. The electric dipole $ed\tau_z$ of the DQD with one electron can be probed via microwave transmission through the cavity. Theoretically, this means that the DQD and cavity need to be treated as an open quantum system. The transmission can be efficiently calculated using input-output theory (see Box 2). It is advantageous to first diagonalize $H_0$ and transform $H_{int}$ into the eigenbasis of $H_0$. Transforming into a frame rotating with the probe field frequency and neglecting fast oscillating terms within the rotating-wave approximation, one finds $H = \frac{\Omega}{2}\tau_z + \tilde{g}_c(a\,\tau_+ + a^\dagger\,\tau_-) + \Delta a^\dagger a$ where $\tilde{g}_c = \frac{g_0\,t_c}{\Omega}$, and where we have added the photon energy in the rotating frame $\Delta/2\pi = f_c - f_R$ with the probe frequency $f_R$ (often $f_c = f_R$ and thus $\Delta = 0$). Here, the $\tau$ operators are defined in the eigenbasis of $H_0$. The Heisenberg-Langevin equations of motion for the photon operator and the electron coherence operator are then found to be (see Box 2)

$$\dot{a} = -i\Delta a - \frac{\kappa}{2}a + \sum_n \sqrt{\kappa_n}\,a_{n,in}(t) - i\tilde{g}_c\tau_-,$$

$$\text{and } \dot{\tau}_- = -i\Omega\tau_- - \frac{\gamma}{2}\tau_- - i\tilde{g}_c a,$$

where we have neglected quantum noise terms [65-67]. Here, in addition to the coherent contributions from the quantum Heisenberg equations of motion, the incoherent terms take into account the cavity decay with rate $\kappa = \kappa_1 + \kappa_2 + \kappa_{int}$ (photon loss at the two ports plus intrinsic losses), the charge qubit decay rate $\gamma$, and the cavity input field $a_{n,in}$ on mirror n. In the stationary limit $\dot{a} = \dot{\tau}_- = 0$ we find for the transmission coefficient through the cavity

$$A = \frac{a_{2,out}}{a_{1,in}} = \frac{\sqrt{\kappa_2}a}{a_{1,in}} = \frac{-i\sqrt{\kappa_1\kappa_2}}{\Delta - \frac{i\kappa}{2} + \tilde{g}_c \chi},$$

with the single-electron electric susceptibility

$$\chi = \frac{\tilde{g}_c}{-\Omega + i\gamma/2}.$$

For simplicity, we can consider a symmetric cavity without intrinsic losses (e.g. $\kappa_{int} = 0$), such that $\kappa_1 = \kappa_2 = \kappa/2$. The cavity is also often probed on resonance ($\Delta = 0$). In the absence of a DQD, $\chi = 0$ and we find unhindered transmission of microwaves through the cavity ($A = 1$). Charge dynamics within the DQD results in an effective microwave admittance that loads the superconducting cavity, changing the cavity amplitude and phase response [68-72]. The electric susceptibility $\chi$ is greatest (and thus $A$ is the smallest) for a symmetric DQD ($\varepsilon = 0$) because in this configuration the electron is most easily transferred from left to right and back.

**Box 2: Input-output theory primer**

The cavity-DQD system can be accurately described using techniques from the theory of open quantum systems[73]. In this formulation we break up the total Hamiltonian $H = H_s + H_r + V_{sr}$ into a system Hamiltonian $H_s$ for the DQD, its surrounding environment, and a single mode of the cavity with bosonic operator $a$, a reservoir Hamiltonian $H_r$ describing a bath of electromagnetic modes $b_n(f)$ for each port ("mirror") coupled to the cavity mode, and an interaction Hamiltonian $V_{sr}$ that couples the cavity mode to the reservoir

$$V_{sr} = \hbar \sum_n \int df\, \lambda_n(f)[b_n(f)\, a^\dagger + a\, b_n^\dagger(f)].$$

Under the condition that the coupling constants $\lambda_n(f)$ are approximately independent of the frequency $f$ over the frequency range of interest, we can treat the reservoir as a Markovian bath. Formally integrating the Heisenberg equation of motion for $b_n(f, t)$ starting from an initial time $t_0 < t$, we arrive at closed Heisenberg equations of motion for $a(t)$

$$\dot{a}(t) = \frac{i}{\hbar}[H_s, a] - \frac{\kappa}{2}a(t) + \sum_n \sqrt{\kappa_n}\, a_{n,in}(t),$$

$$a_{n,in}(t) = \sqrt{2\pi} \int df\, b_n(f, t_0)\, e^{-2\pi i f(t-t_0)},$$

where $\kappa = \kappa_{\text{int}} + \sum_n \kappa_n$ is the total cavity decay rate including intrinsic loss $\kappa_{\text{int}}$, $\kappa_n = 2\pi|\lambda_n(f)|^2$ is the decay rate through port $n$ of the cavity, and $a_{n,in}(t)$ is the "input" field incident on port $n$ of the cavity. Applying a boundary condition on $b_n(f,t)$ at $t_1 > t$ gives rise to a similar equation for $a(t)$ in terms of "output" fields $a_{n,out}(t)$. The input and output fields have the simple relation

$$a_{n,out}(t) = \sqrt{\kappa_n}\, a(t) - a_{n,in}(t),$$

which allows for a complete description of the cavity response. In the main text we consider a two-port system with an input field on port 1 and $a_{2,in}(t) = 0$. The measured transmission coefficient $A$ is then given by the ratio of the output field on port 2 $a_{2,out}(t) = \sqrt{\kappa_2}\, a(t)$ to the input field on port 1 $a_{1,in}(t)$.

**Quantum coherent charge-photon coupling**

The scale of the susceptibility $\chi \propto g_c$ and transmission $A \propto 1/g_c^2$ (assuming $\kappa \ll g_c$ and $\varepsilon = 0$ in the case of a DQD) are both determined by the electric dipole $ed$ and the vacuum cavity electric field $E_0$ via the electron-dipole coupling strength $g_c = eE_0 d$. For a Rydberg atom in an optical cavity, one has $E_0 \approx$ mV/m and $d \approx 100$ nm, leading to couplings $g_c$ roughly in the 10 or 100 kHz range[74]. This coupling can in principle be strengthened in two ways: either by increasing the electric dipole through an increase in size or by increasing the vacuum electric field $E_0 = \sqrt{hf_c/2\epsilon_0 V}$, where $f_c$ denotes the cavity frequency and $V$ the cavity mode volume. While superconducting circuit microwave resonators typically have a slightly lower resonance frequency than three-dimensional cavities for Rydberg atom-based cavity QED, their mode volume can be thousands of times smaller than that of 3D cavities[9, 10]. The vacuum electric field can therefore be several orders of magnitude stronger in superconducting resonators, which allows for qubit-resonator couplings of $g_c/2\pi \approx 1-10$ MHz for quantum dots ($d \approx 100$ nm) and $g_c/2\pi \approx 10-100$ MHz for superconducting qubits ($d \approx 1$ μm). Crucially, such large values of $g_c$ can easily exceed both the cavity linewidth $\kappa$ and qubit decay rate $\gamma$. The limit $g_c \gg \gamma, \kappa$ is called the strong coupling regime of cavity QED. Achieving strong coupling is significant because the qubit and photon degrees of freedom become directly entangled with each other under these conditions[75]. In addition to being of fundamental interest, this entanglement can be exploited for applications in quantum information science[76].

**2. Experimental demonstrations of charge-photon coupling with quantum dots**

Hybrid quantum devices comprising gate-defined quantum dots (QDs) that are coupled via their electric dipole moment to microwave cavities have been successfully demonstrated using multiple material systems including GaAs/AlGaAs heterostructures[35, 77-80], InAs nanowires[67, 81], graphene[82, 83], carbon nanotubes[84, 85] and Si/SiGe heterostructures[34, 36, 37, 86]. The microwave cavity is often realized as a superconducting coplanar waveguide resonator[9, 10, 14, 24, 87], with an example shown in the top left panel of FIG. 3a. In order to maximize the quality factor of the cavity and the chance of reaching the strong-coupling regime, each gate line leading to the DQD is sometimes filtered by an on-chip low pass $LC$-filter to suppress photon leakage from the cavity[86].

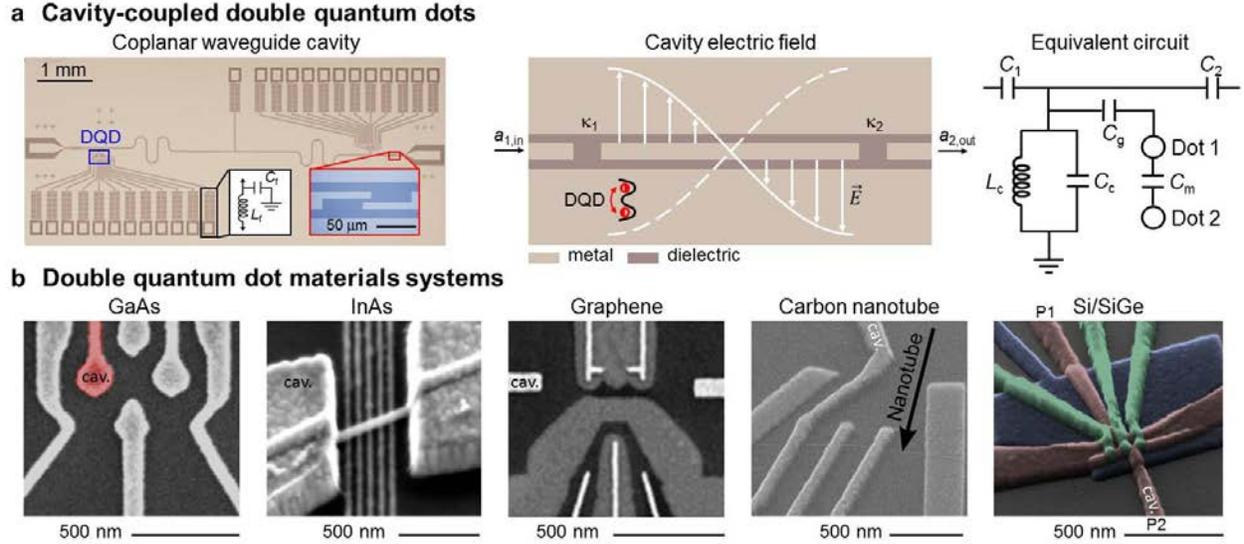

Figure 3 | **Constructing cavity-coupled double quantum dots. a** | Left panel: Optical image of a Nb coplanar waveguide cavity fabricated on top of a Si/SiGe heterostructure (adapted with permission from REF.[34], AAAS). The coplanar waveguide cavity is located in the middle of the sample and is coupled to measurement ports through vacuum gap capacitors located at each end of the cavity (see right inset). A gate-defined DQD is positioned at an anti-node of the cavity electric field. Low-pass *LC*-filters, as represented by the circuit diagram in the left inset, allow for dc biasing of the DQD gate electrodes and reduce photon losses from the cavity. Middle panel: Schematic representation of the device. A half-wavelength ($\lambda/2$) standing wave is formed in the coplanar waveguide cavity (white lines) and couples to a single electron trapped in the DQD via the electric-dipole interaction. In a typical experiment, port 1 of the cavity is driven by a coherent microwave field $a_{1,\text{in}}$ and the signal exiting port 2 of the cavity $a_{2,\text{out}}$ is measured. $\kappa_1$ ($\kappa_2$) denotes the coupling rate between the cavity and port 1 (port 2). Right panel: Circuit representation of the device. Here the microwave cavity is modeled as a parallel *LC*-oscillator with an effective inductance $L_c$ and capacitance $C_c$. The DQD is modeled as a pair of charge islands with a mutual capacitance $C_m$ and dot 1 is coupled to the *LC*-oscillator through a capacitance $C_g$. $\kappa_1$ ($\kappa_2$) is set by the port capacitance $C_1$ ($C_2$). **b** | Scanning electron micrographs of cavity-coupled DQDs fabricated from a variety of host materials, including GaAs (adapted with permission from REF.[88]), InAs (adapted with permission from REF.[89], American Physical Society), graphene (adapted with permission from REF.[83], American Chemical Society), carbon nanotubes (adapted with permission from REF.[90], American Institute of Physics) and Si/SiGe (adapted with permission from REF.[34], AAAS). The gate electrode connected to the microwave cavity is indicated for each device by the letters "cav.". The gates labeled $V_{P1}$ (left dot) and $V_{P2}$ (right dot) in the Si/SiGe device are used to adjust the DQD level detuning ε.

A simplified schematic of the hybrid device is depicted in the top middle panel of FIG. 3a. At the fundamental resonance frequency $f_c$, the vacuum fluctuation of the cavity generates a half-wavelength $\lambda/2$ electromagnetic standing wave[10, 91]. At a voltage anti-node of the standing wave, a delocalized electron occupying the molecular bonding and anti-bonding states of the DQD[47, 48, 60] couples to the electric field of the cavity via the electric-dipole interaction[67, 77]. A circuit representation of the device is shown in the right panel of FIG. 3a. Here the cavity is modeled as a parallel *LC*-oscillator having an effective inductance $L_c$, effective capacitance $C_c$ and resonance frequency $f_c = \sqrt{1/L_c C_c}/2\pi$ (REFS[92, 93]). The DQD is mutually coupled via a capacitance $C_m$ and dot 2 is capacitively coupled to the cavity via $C_g$. The system is connected to an input port via capacitor $C_1$ and an output port via capacitor $C_2$, allowing for measurements of the cavity transmission amplitude $A = |a_{2,out}/a_{1,in}|$ and phase $\phi = -\arg(a_{2,out}/a_{1,in})$ using homodyne or heterodyne detection techniques[10, 20, 94].

Since the charge-photon coupling rate $g_c$ scales linearly with $\sqrt{Z_c}$, where $Z_c = \sqrt{L_c/C_c}$ is the characteristic impedance of the cavity[9, 55], it is desirable to increase $Z_c$ beyond the range between 20 Ω and 200 Ω that is the typical limit of co-planar waveguide cavities[93]. One way of increasing the impedance is to define the microwave cavity using a linear array of superconducting quantum interference devices (SQUIDs) made from Al Josephson junctions, leading to $L_c \approx 1.5$ kΩ by virtue of the large Josephson inductance of each SQUID[35]. Another approach, discussed in the next section, utilizes the large kinetic inductance of a nanowire made from NbTiN[37, 95].

Detailed scanning electron micrographs (SEMs) of cavity-coupled DQDs made with different host materials are shown in FIG. 3b. In the case of GaAs or Si, one or three layers of surface gate electrodes are directly patterned on top of the buried quantum well (QW) of a GaAs/AlGaAs or Si/SiGe heterostructure[77, 86]. For InAs nanowires, graphene and carbon nanotubes, the host material is first transferred to a Si substrate before the patterning of gate electrodes[67, 82, 90]. To maximize $g_c$, an electrode is often galvanically connected to the center pin of the superconducting cavity[34, 67, 77, 78, 82].

**Strong charge-photon coupling**

Achieving the strong-coupling regime for DQD charge qubits is generally challenging due to their typically large decoherence rates $\gamma_c$, which commonly fall between a few hundred MHz and several GHz in earlier works[47, 49, 63, 67, 77-79, 82, 85, 96, 97]. These values often exceed the coherent charge-photon coupling rate $g_c$ by one or more orders of magnitude[67, 77-80, 82, 85, 96, 97]. Therefore, a significant reduction in $\gamma_c$ or a significant increase in $g_c$ is needed to access the strong-coupling regime $g_c > (\gamma_c, \kappa)$. Both approaches have recently been met with success in two experiments which we review here[34, 35].

A first step toward charge-photon coupling is the detection of charge states within the DQD. This is traditionally accomplished by measuring the conductance of a proximal quantum point contact (QPC) which is sensitive to the charge distribution within the QDs[98-100]. Charge state detection may also be performed by measuring the transmission properties of the cavity, which are sensitive to the tunnel-rate-dependent complex admittance of the QDs[68, 69, 72]. An example is shown in the upper panel of FIG. 4a. Here the cavity transmission amplitude $A/A_0$ ($A_0$ is a normalization constant) at a fixed drive frequency $f = f_c$ is measured as a function of gate voltages $V_L$ and $V_R$ (e.g., see right-most bottom panel of FIG. 3b), which control the chemical potentials of the left and right dot, respectively. The charge stability diagram characteristic of a few-electron DQD is clearly visible[60], where we have used ($N_L$,$N_R$) to denote a charge state with $N_L$ electrons in the left dot and $N_R$ electrons in the right dot. To determine $g_c$, $A/A_0$ is measured around an interdot charge transition ($N_L$+1,$N_R$)−($N_L$,$N_R$+1) (bottom panel of FIG. 4a, taken with $2t_c/h < f_c$)[60]. Here a pair of minima are observed along the DQD detuning axis $\varepsilon$ at locations where $\Omega/h = f_c$. At these detunings the charge qubit is strongly hybridized with the cavity photons[34, 67, 77]. Detailed fitting of the response $A(\varepsilon)/A_0$ to input-output theory allows the charge-photon coupling rate $g_c$ to be extracted from this measurement[34, 67, 77].

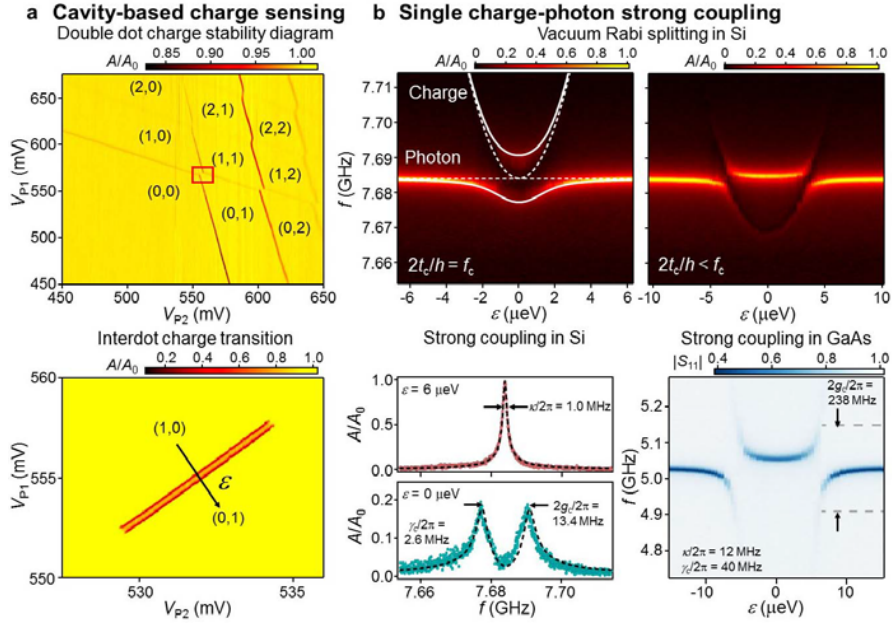

Figure 4 | **Strong charge-photon coupling. a** | Upper panel: The charge stability diagram of the DQD can be extracted by measuring the cavity transmission amplitude $A/A_0$ as a function of the left and right gate voltages. Dot-lead transitions correspond to the exchange of an electron between the source/drain reservoirs and the DQD. Interdot charge transitions correspond to the transfer of an electron between dots and are also visible in the data. Bottom panel: Cavity transmission amplitude $A/A_0$ measured in the vicinity of the (1,0)−(0,1) interdot transition of the same device after the interdot tunnel coupling $t_c$ has been increased. The arrow denotes the axis along which the DQD level detuning parameter $\varepsilon$ is defined. **b** | Top panels: Cavity transmission amplitude $A/A_0$ of a Si-based device as a function of $\varepsilon$ and the drive frequency $f$, taken with $2t_c/h = f_c$ (top left panel) and $2t_c/h < f_c$ (top right panel). The white solid (dashed) lines indicate the eigenenergies of the Jaynes-Cummings Hamiltonian describing the charge-photon system, when $g_c/2\pi = 6.7$ MHz (0 MHz). Bottom left panels: $A/A_0$ as a function of $f$ when $\varepsilon = 6$ μeV and $\varepsilon = 0$. Dashed lines are fits to cavity input-output theory. Bottom right panel: Cavity reflectance spectrum $|S_{11}|$ of a GaAs-based device taken with $2t_c/h < f_c$. Data in the top panels and bottom left panel of panel **b** are adapted with permission from REF.[34], AAAS. Data in the bottom right panel of panel **b** are adapted with permission from REF.[35], American Physical Society.

The hallmark of the strong-coupling regime is the vacuum Rabi splitting, which is the emergence of a pair of distinct resonance peaks in the cavity transmission spectrum for a fixed detuning $\varepsilon$ where the qubit and a cavity photon become equal in frequency[6, 7, 10, 41]. The top left panel of FIG. 4b shows a measurement of $A/A_0$ as a function of $f$ and $\varepsilon$, taken with a Si-based DQD tuned to $2t_c/h = f_c$ (REF[34]). At a large value of $\varepsilon = 6$ μeV, the cavity transmission spectrum (bottom left panel of FIG. 4b) exhibits a single peak. Here the qubit-photon frequency detuning is large (white dashed lines) and the full-width-at-half-maximum (FWHM) of $A^2/A_0^2$ gives the bare cavity loss rate $\kappa/2\pi = 1.0$ MHz. At $\varepsilon = 0$, the DQD splitting $\Omega = 2t_c = hf_c$ and the charge qubit become entangled with a single photon, leading to an "avoided crossing" in the eigenenergies of the charge-photon system (white solid lines). Correspondingly, vacuum Rabi splitting is observed in the cavity transmission spectrum (bottom left panel of FIG. 4b) where the two resonance modes are separated by a vacuum Rabi frequency $2g_c/2\pi = 13.4$ MHz. The top right panel of FIG. 4b shows similar data taken with $2t_c/h < f_c$, in which a pair of avoided crossings are observed when $\Omega/h = f_c$ at finite $\varepsilon$. The clear resolution of the vacuum Rabi splitting suggests that the regime of strong charge-photon coupling has been achieved for this Si-based device – a conclusion further supported by a charge decoherence rate $\gamma_c/2\pi = 2.6$ MHz independently determined using microwave spectroscopy in the dispersive regime[34, 101]. The charge decoherence rate of this device, about two to three orders of magnitude lower than typical DQD charge qubits[47, 49, 63, 67, 77-80, 82, 85], is a subject of ongoing investigation. A more recent work suggests that it may arise from both a low level of charge noise and the effect of valley-orbit hybridization[102].

Strong charge-photon coupling has also been demonstrated with GaAs-based devices, using a SQUID array cavity[35]. Due to the higher impedance of the microwave cavity, a large charge-photon coupling rate $g_c/2\pi = 119$ MHz is found (bottom right panel of FIG. 4b), which allows the strong-coupling regime to be accessed despite comparatively large values of $\gamma_c/2\pi = 40$ MHz and $\kappa/2\pi = 12$ MHz. More recently, a low charge decoherence rate of $\gamma_c/2\pi = 3$ MHz has been achieved by a GaA-based DQD as well[88].

### 3. Spin-photon coupling

The quantum coherence of the spin ½ of individual electrons in quantum dots or defects in silicon typically lasts between tens of microseconds to several milliseconds[103-106] and can in some cases even approach a second[107], while the nuclear spin coherence can last as long as a minute[107]. In comparison, the coherence of the charge qubit in a DQD is quite short-lived with a decay time of typically a few nanoseconds [47-49]. The spin is therefore the primary choice as a qubit for quantum information processing in semiconductors[108, 109]. Since the exchange interaction is short ranged[51], this naturally leads to the question of how to couple two electron spins that are separated by a large distance using spin-electric coupling to a common cavity mode. At the face of it, this seems very hard because the spin of an electron does not directly couple to the electric field of the cavity. However, there are several techniques to hybridize the spin and charge degrees of freedom (qubits) of an electron. All of these methods endow the spin with an effective electric dipole that enables its interaction with the electric field of the cavity. For multi-electron spin qubits, the Fermi statistics provides a way to couple orbital and spin degrees of freedom[12, 110, 111] (a recent experiment using this method has attained strong spin qubit-photon coupling[38]). One mechanism that works for single electron spins is the natural built-in spin-orbit coupling due to relativistic effects which can be sizeable in a number of semiconductor materials[13, 112, 113]. The intrinsic spin-orbit coupling may work particularly well for holes in the valence band of some semiconductors[114]. Without relying on such intrinsic effects, one can engineer a spin-electric interaction using controlled magnetic fields, either time-dependent fields that induce electron spin resonance[11, 115-117] or static but spatially varying fields produced by an on-chip microscale ferromagnet[36, 64, 105, 118, 119]. In the case of a static magnetic field gradient $\nabla_x \boldsymbol{B}$ produced by a micromagnet, an applied electric field $E_{ac}$ will shift the electron position in a single quantum dot by $x_E = eE_{ac}a_0^2/E_{orb}$ where $E_{orb}$ and $a_0$ denote the energy level spacing and size of the quantum dot. For an oscillatory electric field this means that the magnetic field seen by the electron also becomes oscillatory, $\boldsymbol{B}(x_0 + x_E \sin \omega t) \sim \boldsymbol{B}(x_0) + \nabla_x \boldsymbol{B} \, x_E \sin \omega t$, allowing for electric dipole spin resonance (EDSR). For the quantized cavity field, one finds that $E_{ac} \sin \omega t$ is replaced by $E_{cav} = E_0(a + a^\dagger)$ resulting in a spin-photon coupling $g_s \sim eE_0 \nabla_x \boldsymbol{B} \, a_0^2/E_{orb}$. The spin-phonon coupling $g_s$ in a DQD can be much larger and more controllable than for a single QD. For a symmetric DQD at $\varepsilon = 0$ one finds $g_s \sim eE_0 \nabla_x \boldsymbol{B} \, d^2/\Omega \sim g_c \Delta B_x / \Omega$, where $d$ is the distance between the two dots, $\Delta B_x = \nabla_x \boldsymbol{B} \, d$ is the change in magnetic field (measured in energy units) from one dot to the other, and the DQD energy splitting $\Omega$ can be tuned by the inter-dot tunnel coupling and the external magnetic field[64]. Since $d > a_0$ and $\Omega \ll E_{orb}$, $g_s$ is much larger in a DQD compared to a single dot [120, 121].

To study the combined charge and spin dynamics of the spin and charge of a single electron in a DQD one can employ the 4x4 Hamiltonian in the basis $|(\uparrow,0)\rangle, |(\downarrow,0)\rangle, |(0,\uparrow)\rangle, |(0,\downarrow)\rangle$

$$H_0 = \frac{1}{2}\begin{pmatrix} \varepsilon + B_z & \Delta B_x & 2t_c & 0 \\ \Delta B_x & \varepsilon - B_z & 0 & 2t_c \\ 2t_c & 0 & -\varepsilon + B_z & -\Delta B_x \\ 0 & 2t_c & -\Delta B_x & -\varepsilon - B_z \end{pmatrix},$$

which includes the Zeeman coupling $H_Z = \boldsymbol{S} \cdot \boldsymbol{B}(\mathbf{r})$ of the spin $\boldsymbol{S}$ to an external magnetic field $\boldsymbol{B}(\mathbf{r})$ (in energy units) [REF[64]]. A magnetic field $B_z$ pointing in z direction leads to an energy splitting between the spin-up and spin-down states. As long as the field has the same strength in both dots, i.e. $\boldsymbol{B}$ does not depend on the position $\mathbf{r}$, the spin and charge qubits are completely separate. In this case, only the charge qubit interacts with the electromagnetic field (photons) of the cavity, while the spin is decoupled from it. However, as soon as a magnetic field difference $\Delta B_x$ perpendicular to the homogeneous field component is applied, the charge and spin qubits are hybridized, allowing for a coupling of the spin qubit to the cavity photons. The coupling to the cavity is again obtained by replacing $\varepsilon$ with $\varepsilon + eE_{cav}d$ with the cavity electric field $E_{cav} = E_0(a + a^\dagger)$. In this way, we obtain the Hamiltonian $H = H_0 + H_{int}$ with $H_{int} = g_c(a + a^\dagger)\tau_z$ where $g_c = eE_0 d$. The four relevant energy levels $|n\rangle$ of the DQD are found by diagonalizing the matrix $H_0$, while the electric dipole transition matrix elements $d_{nm}$ can be determined by transforming $H_{int}$ into the eigenbasis of $H_0$,

$$H_{\text{int}} = g_c(a + a^\dagger) \sum_{n,m=0}^{3} d_{nm} |n\rangle\langle m|.$$

For the understanding of the most important mechanisms for the spin-photon interaction, it is sufficient to consider an effective two-level model. Making also the rotating wave approximation, one arrives at the Jaynes-Cummings model $H = \frac{1}{2}\Delta_s \sigma_z + g_s(a\sigma_+ + a^\dagger \sigma_-)$, where the $\sigma_z$ Pauli operators act on the low-energy hybridized spin states, and $\Delta_s = B_z - hf_c$ is the detuning of the spin splitting $B_z$ from the photon energy $hf_c$. For a symmetric DQD with $\varepsilon = 0$ one finds a spin-photon coupling rate $g_s \cong g_c \Delta B_x/(2\,\delta)$ where $\delta = 2t_c - hf_c$ can be controlled by adjusting the tunnel coupling $t_c$ between the two quantum dots. While this two-level model explains the vacuum Rabi splitting that has been observed experimentally, there are more subtle effects such as the asymmetry of the Rabi peak heights that require a three-level model for their explanation[64].

**Strong spin-photon coupling**

Compared to charge-photon coupling, reaching the strong-coupling regime of spin-photon interaction faces a distinct challenge: The direct magnetic-dipole coupling rate $g_s$ between a single electron spin and a single photon is mostly limited to between 10 Hz and 500 Hz, which is too slow to overcome single-spin dephasing rates or cavity loss rates[44, 45, 122-125]. As such, a robust scheme for spin-charge hybridization is necessary to increase $g_s$ to the MHz range where strong-coupling becomes feasible[11-13, 64, 110, 111, 120, 121, 126-128]. At the same time, a low level of charge noise is required of the device since spin-charge hybridization subjects the electron spin to charge-noise-induced dephasing. A cavity-coupled carbon nanotube DQD in an earlier experiment hybridized spin and charge via ferromagnetic leads to achieve $g_s/2\pi = 1.3$ MHz, but was still in the weak-coupling regime due to a larger spin decoherence rate $\gamma_s/2\pi = 2.5$ MHz (REF[129]). More recently, two experiments using Si-based DQDs have successfully attained the strong-coupling regime between a single spin and a single photon[36, 37]. We review these results in this section.

The setup for one of the experiments[36] is illustrated in FIG. 5a. The device is a gate-defined DQD on top of a Si/SiGe heterostructure coupled to a co-planar waveguide cavity, similar to the previous work on strong charge-photon coupling[34] but including a crucial new ingredient: the addition of a micron-sized Co magnet on top[103, 130, 131]. When magnetized by an externally applied magnetic field $B_z^{\text{ext}}$, the fringing field of the micromagnet creates a large gradient for the magnetic field component pointing along the x-axis, i.e., a large $\partial B_x/\partial z$. As such, the quantization axis of the electron spin is dependent on its location, hybridizing the spin and charge degrees of freedom[13, 64, 120, 121, 128]. A plot of the DQD energy levels including the spin degree of freedom is provided in FIG. 5b, which is helpful for discussing the experimental results below.

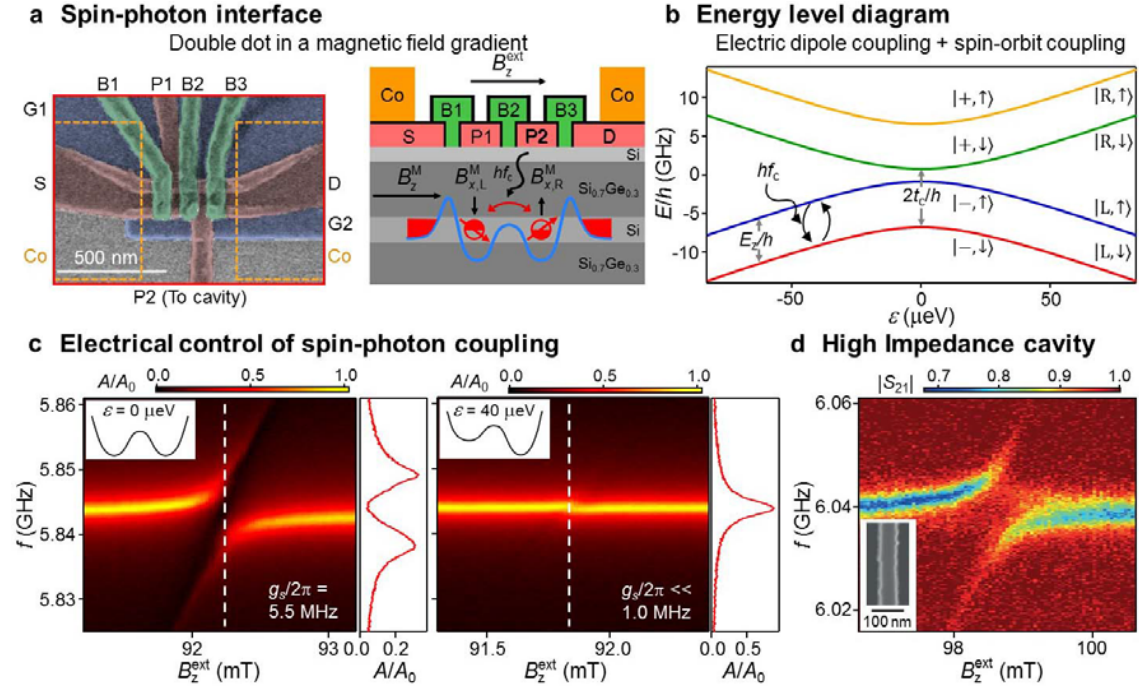

Figure 5 | **Strong spin-photon coupling. a** | Top left panel: SEM image of a Si/SiGe DQD used to achieve spin-photon coupling. The orange dashed lines represent the locations of a pair of Co micromagnets fabricated on top of the DQD. Top right panel shows a cross-sectional view of the device. The application of an external magnetic field $B_z^{\text{ext}}$ polarizes the micromagnets and creates an inhomogeneous magnetic field having a component $B_z^{\text{M}}$ parallel to $B_z^{\text{ext}}$ and a component $B_x^{\text{M}}$ orthogonal to $B_z^{\text{ext}}$. $B_x^{\text{M}}$ changes sign between the two dots, assuming a value of $B_{x,\text{L}}^{\text{M}}$ for the left dot and $B_{x,\text{R}}^{\text{M}}$ for the right dot. As a result, the quantization axis of an electron's spin (red arrows) is dependent on the electron's position. **b** | Energy level diagram of a single electron trapped in a DQD in the presence of an inhomogeneous magnetic field as a function of the DQD detuning energy ε. Here ↑ and ↓ denote the Zeeman-split spin-states of the electron, $L$ ($R$) denotes the single-dot orbital state of the left (right) dot and – (+) denotes the molecular bonding (anti-bonding) state formed by the hybridization of the $L$ and $R$ states. **c** | Left panel: Cavity transmission amplitude $A/A_0$ as a function of $f$ and $B_z^{\text{ext}}$. Vacuum Rabi splitting with a frequency $2g_s/2\pi = 11.0$ MHz is observed at $B_z^{\text{ext}} = 92.2$ mT. Right panel: Increasing the detuning to ε = 40 μeV greatly reduces the vacuum Rabi splitting, allowing for electrical control of the spin-photon coupling rate. **d** | Strong spin-photon coupling using a high impedance NbTiN nanowire resonator. The cavity transmission coefficient $|S_{21}|$ is plotted as a function of $f$ and $B_z^{\text{ext}}$. Inset: Scanning electron microscope image of a portion of the NbTiN nanowire resonator. Panels **a**, **b** and **c** are adapted from REF.[36], Macmillan Publishers Limited. Panel **d** is adapted with permission from REF.[37], AAAS.

To search for spin-photon coupling, the frequency of the single-spin qubit $E_Z/h = g\mu_B B_{tot}/h$ is tuned into resonance with the cavity by changing the external magnetic field $B_z^{ext}$. Here $E_Z$ is the Zeeman energy, $g$ is the g-factor of the electron and $\mu_B$ is the Bohr magneton. $B_{tot}$ is the total magnetic field spatially averaged over the electron's wavefunction, having contributions from both the externally applied field and the intrinsic field of the micromagnet. The left panel of FIG. 5c shows the cavity transmission amplitude $A/A_0$ as a function of $f$ and $B_z^{ext}$, taken at $\varepsilon = 0$. A clear avoided crossing is observed around $B_z^{ext} = 92.2$ mT, where the resonance condition $E_Z/h = f_c$ is met. A plot of $A/A_0$ as a function of $f$ at $B_z^{ext} = 92.2$ mT again shows vacuum Rabi splitting with a frequency $2g_s/2\pi = 11.0$ MHz (left panel of FIG. 5c), signifying strong-coupling between the single electron spin and a cavity photon. The spin-photon coupling rate $g_s/2\pi = 5.5$ MHz observed here exceeds direct magnetic-dipole coupling rates by four to five orders of magnitude[44, 45, 122-125]. This remarkable enhancement in $g_s$ may be understood by considering the energy diagram in FIG. 5b. In the regime $|\varepsilon| \ll t_c$, the single-dot orbital states $L$ and $R$ are hybridized by the interdot tunnel coupling $t_c$ to form molecular bonding and anti-bonding states (see discussion in Box 1). While occupying these charge states, the electron wavefunction becomes delocalized across the two dots and the electric field of a cavity photon can displace its wavefunction by about 1 nm, generating a large effective magnetic field due to the magnetic field gradient and yielding a large $g_s$ (REF[36]).

A second experiment, also involving a DQD defined on a Si/SiGe heterostructure, uses a cavity design composed of a thin NbTiN nanowire with a large kinetic inductance (inset to FIG. 5d)[37]. A higher impedance in the kilo-ohm range is supported by this cavity. Strong spin-photon coupling is also achieved by this device, as shown by the avoided crossing in FIG. 5d.

To apply the spin-photon cavity QED device to quantum information processing, it is also necessary to rapidly switch on and off the spin-photon coupling rate $g_s$. This flexibility would allow the spin qubit to be manipulated in an isolated state ($g_s \approx 0$) where it is protected from cavity-induced Purcell decay[123, 132-134] and read out via the cavity when the coupling is back on. One way to tune $g_s$ is by tilting the DQD potential, as shown in FIG. 5c. As $\varepsilon$ is increased from zero, we observe a strong decrease of spin-photon coupling from $g_s/2\pi = 5.5$ MHz ($\varepsilon = 0$) to $g_s/2\pi \ll 1$ MHz ($\varepsilon = 40$ μeV). This change is due to the fact that at $|\varepsilon| \gg t_c$, the electron wavefunction becomes strongly localized within one dot (FIG. 5b) and interdot tunneling is largely suppressed. Here the displacement of the electron wavefunction by the cavity photon is limited to about 3 pm in distance[36]. The effective magnetic field generated by a cavity photon is therefore very small, effectively turning off spin-photon coupling. Using nanosecond control of $\varepsilon$, driven Rabi oscillation and dispersive readout of the single-spin qubit have been demonstrated[36], paving the way toward quantum non-demolition readout of spin qubits[135] which may allow error-correction codes such as the surface code to be implemented with spin qubits[136, 137].

## 4. Outlook/Conclusions

Where do these exciting developments lead us? The strong and controllable coupling between individual spin qubits embedded in a superconducting microwave resonator allows for long-distance spin-spin coupling mediated by microwave photons[138, 139]. This coupling can then be employed to perform entangling two-qubit gates between spins separated by several millimeters. One should keep in mind that one millimeter is a very long distance compared with the 80 nm separation of nearest-neighbor spin qubits in Si[140]. Their small footprint on a semiconductor chip is one characteristic feature of semiconductor spin qubits which makes them strong contenders for a scalable quantum information processing platform. In addition to providing the possibility to entangle distant spin qubits, non-local two-qubit gates may facilitate quantum error correction in the framework of a fault-tolerant quantum computing architecture. Also in this context, the possibility of creating a network of spin qubits with engineered coupling may be very useful for realizing a surface code[136]. Moreover, the possibility of creating a network of spin qubits with coupling geometries ranging from local to "all-to-all" opens interesting perspectives for quantum simulation of interacting quantum many-body systems[141, 142].

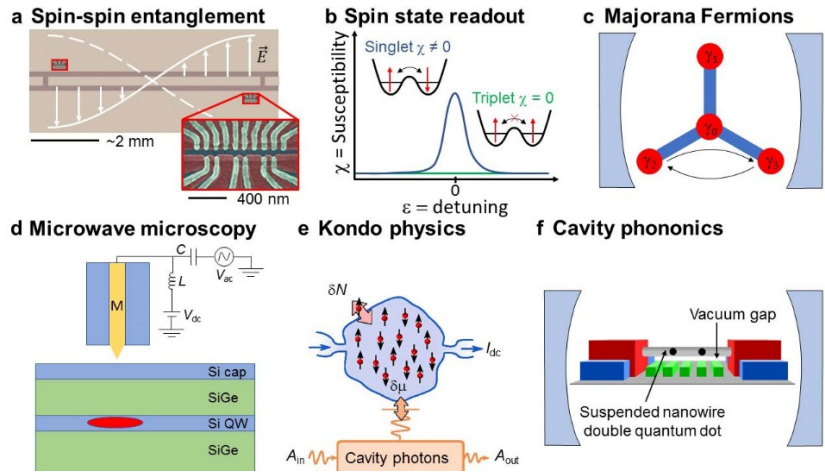

Figure 6 | **Future directions in "super-semi" circuit QED. a** | A future spin-based quantum processor could consist of local nodes of nearest-neighbor exchange coupled spins. Long distance coupling of the nodes could be achieved using spin-photon coupling. **b** | The charge susceptibility of a two-electron DQD is dependent on the electron spin configuration, leading quite naturally to cavity-based readout of electron spin states. **c** | Circuit QED has been proposed as a platform to allow for braiding of Majorana Fermions. **d** | Scanning microwave impedance microscope based on circuit QED. In this example the evanescent field from a scannable superconducting resonator is used to probe the valley splitting in a quantum dot that is induced beneath the scanning probe. **e** | Circuit QED may be used to probe Kondo physics in carbon nanotubes. The measurement technique may also shed light on other exotic states of matter. **f** | The transmission through a superconducting cavity has been shown to be sensitive to mechanical degrees of freedom. Coherent conversion from optical to microwave frequencies has been attempted using circuit QED devices incorporating mechanical resonators.

Spin-photon coupling has important implications beyond the generation of long-range quantum entanglement (FIG. 6a). The coupling of the electron spin to an electromagnetic cavity also allows for the dispersive readout of the quantum state of the spin qubit[36, 67] and lays the groundwork for the development of quantum non-demolition[135, 143] and single-shot readout methods[144]. Since the spin-photon coupling gate $g_s$ is a strong function of detuning ε, electrically switching on the cavity coupling of each spin qubit[34] may allow for selective readout in large arrays of spin qubits (FIG. 6b). Moreover, the superconducting qubit community has adopted the use of frequency multiplexed resonators[21, 145] for quantum state readout. A similar approach could be adopted for spins[146].

Looking well beyond spin qubits, the nascent field of hybrid circuit quantum electrodynamics could have a major impact on condensed matter physics as a whole. Some potential areas of research are illustrated in FIG. 6. Cavity measurements have been proposed to investigate Majorana modes[147] and provide an alternative to the somewhat ambiguous measurements of zero-bias conductance peaks[148-151]. It has even been suggested that microwave cavities could be used to implement braiding of Majorana Fermions[152] (FIG. 6c). Scanning-probe versions of superconducting cavities (FIG. 6d) could be used to probe valley physics in silicon[153-156] and perhaps be of much broader use in investigations of two-dimensional quantum materials[157]. Lastly, superconducting cavities have been shown to provide an alternative means to investigate Kondo physics[158-160] (FIG. 6e) and electron-phonon coupling[161, 162] (FIG. 6f). Clearly these applications are just scratching the surface and there are many unopened areas of investigation, including, for example, spin-charge separation in Luttinger liquids[163-165] and THz probes of topological phases of matter[166].